\documentclass[3p,times, twocolumn]{elsarticle}

\usepackage{lineno,hyperref}
\modulolinenumbers[5]

\makeatletter
\def\ps@pprintTitle{%
  \let\@oddhead\@empty
  \let\@evenhead\@empty
  \def\@oddfoot{\reset@font\hfil\thepage\hfil}
  \let\@evenfoot\@oddfoot
}
\makeatother

\begin{document}

\begin{frontmatter}

\title{Structural phase transitions and topological defects in ion Coulomb crystals}

\author{Heather L. Partner\fnref{PTB}}
\author{Ramil Nigmatullin\fnref{Ulm1}}
\author{Tobias Burgermeister\fnref{PTB}}
\author{Jonas Keller\fnref{PTB}}
\author{Karsten Pyka\fnref{PTB}}


\author{Martin B. Plenio\fnref{Ulm2,Ulm3}}
\author{Alex Retzker\fnref{Racah}}


\author{Wojciech H. Zurek\fnref{LANL}}
\author{Adolfo del Campo\fnref{UMB}}


\author{Tanja E. Mehlst\"aubler\fnref{PTB} \corref{mycorrespondingauthor}}


%

\cortext[mycorrespondingauthor]{email: tanja.mehlstaeubler@ptb.de}

\address[PTB]{Physikalisch-Technische Bundesanstalt, Bundesallee 100, 38116 Braunschweig, Germany}
\address[Ulm1]{Institute of Quantum Physics, Albert Einstein Allee 11, Ulm University, 89069 Ulm, Germany}
\address[Ulm2]{Center for Integrated Quantum Science and Technology, Ulm University, Albert Einstein Allee 11, Ulm University, 89069 Ulm, Germany}
\address[Ulm3]{Institute for Theoretical Physics, Albert-Einstein-Allee 11, Ulm University, 89069 Ulm, Germany}
\address[Racah]{Racah Institute of Physics, The Hebrew University of Jerusalem, Jerusalem 91904, Givat Ram, Israel}
\address[LANL]{Theoretical Division, Los Alamos National Laboratory, Los Alamos, NM 87544, USA}
\address[UMB]{Department of Physics, University of Massachusetts Boston, Boston, MA 02125, USA}

\begin{abstract}

We use laser-cooled ion Coulomb crystals in the well-controlled environment of a harmonic radiofrequency ion trap to investigate phase transitions and defect formation. Topological defects in ion Coulomb crystals (kinks) have been recently proposed for studies of nonlinear physics with solitons and as carriers of quantum information. Defects form when a symmetry breaking phase transition is crossed nonadiabatically. For a second order phase transition, the Kibble-Zurek mechanism predicts that the formation of these defects follows a power law scaling in the rate of the transition.  We demonstrate a scaling of defect density and describe kink dynamics and stability. We further discuss the implementation of mass defects and electric fields as first steps toward controlled kink preparation and manipulation.
\end{abstract}


\end{frontmatter}



\section{Introduction}

The invention of the Paul trap in the 1950s \cite{RevModPhys.62.531} enabled the controlled isolation of single ions or small crystals of ions from their environment in a pseudopotential created by oscillating electric fields. 
 With the development of laser cooling, these individually stored ions can be Doppler cooled to minimum temperatures below 1 mK, while more sophisticated cooling techniques can even bring the ions to the quantum mechanical ground state \cite{Metcalf1999, Wineland1979}.  With these advances, individual trapped ions became a well-controlled system that provided a very promising candidate for quantum computation \cite{Steane1997}, as well as quantum simulation of solid state and other systems \cite{Schneider2012}.  

When multiple ions are trapped and laser cooled to energies well below the Coulomb potential energy, crystals form, known as Coulomb crystals \cite{Diedrich1987}. 
 This is analogous to the formation of Wigner crystals in ultra-cold electron gases \cite{Wigner1934, Grimes1979,Chaplik1980, Mehta2013}. 
 Ion Coulomb crystals have been studied theoretically and experimentally since the 1980s \cite{Diedrich1987,Dubin1999, Wineland1987, Piacente2004}, and recently groups have started to investigate their thermodynamic properties.  Fishman et al. \cite{Fishman2008} studied the phonon modes in one- and two-dimensional crystals, and showed that the transition from the linear to zigzag phase in Coulomb crystals is of second-order, identifying the order parameter and deriving the critical exponents using the Ginzburg-Landau model.  It was suggested in \cite{Retzker2008} and described theoretically in \cite{delCampo2010,deChiara2010} that topological defects could be created while crossing this phase transition.

Triggered by the prospect of using laser-cooled ions as a system for quantum computation \cite{Kielpinski2002, Monroe1995}, trap technology has experienced great advancements in the last years. These advances have allowed trapping of large crystals in a highly controlled way  \cite{Blatt2012}.  Such crystals can be manipulated by varying trap parameters, implementing electric and magnetic fields, and performing laser spectroscopy. Individual ions can be observed using fluorescence techniques combined with spatially resolved camera imaging.  This opens up ion Coulomb crystals as a test platform for non-equilibrium statistical mechanics, e.g. to study phase transitions, thermalization, friction, etc. When controlled sufficiently well (i.e. with minimized micromotion) such crystals can be used as a high-stability atomic frequency standard \cite{Pyka2013b,Herschbach2012}. With this evolved experimental system, it has recently become possible to create topological defects in the laboratory \cite{Pyka2013a,Mielenz2013,Ulm2013,Ejtemaee2013}.  Such defects have been proposed for storage of quantum information, and for studies of the nonlinear physics of solitons and phase transitions \cite{Retzker2008,deChiara2010,Nigmatullin2011,PhysRevB.40.2284,Dantan2009}.

In this paper we present experiments in which we trap up to 50 ions and drive nonadiabatic quenches across the phase transition from linear to zigzag.  We discuss the creation, properties, and behavior of different types of topological defects that arise during this transition.  Finally, we describe ways to control and manipulate these kink solitons using varied trap parameters, mass impurities, and electric fields.

\begin{figure}[b]
   \centering
  \includegraphics[scale=0.383]{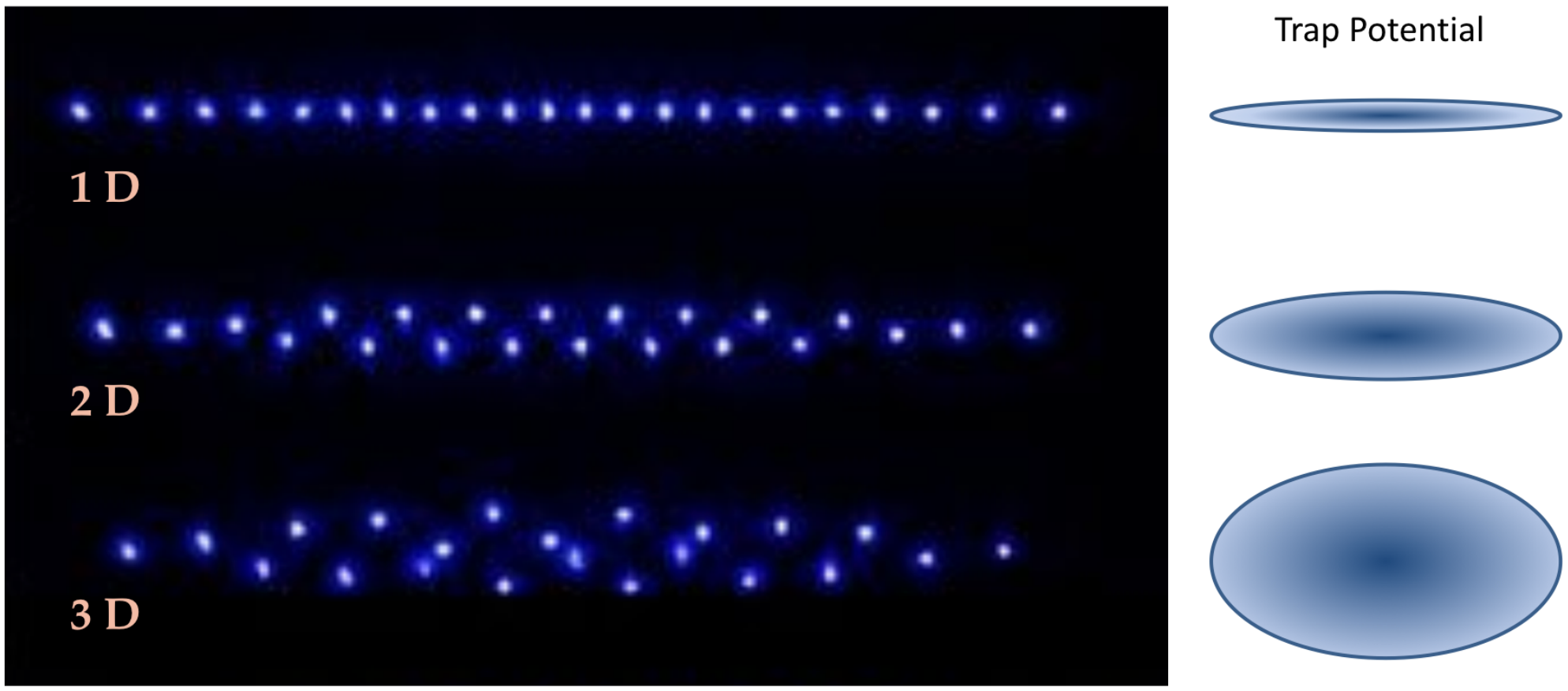} 
   \caption{Three phases (linear, zigzag, helix) of ion Coulomb crystals in our experiment.  The inhomogeneous nature of the axial confinement can be seen in the 2D and 3D crystals where the central ions undergo the transition to the next phase before the rest of the crystal.}
	\label{fig:phases}
\end{figure}

\section{Topological defect creation and dynamics}

Our experimental system consists of a linear Paul trap in which the radial confinement of the linear crystal can be varied by changing the RF trap voltage, and the axial confinement can be varied by adjusting the DC voltages on the endcap electrodes. We image the ions directly onto an Electron Multiplying Charge-Coupled Device (EMCCD) camera chip using a home-built lens system, and with a beam-splitter we optionally also send fluorescence to a Photo-Multiplier Tube (PMT).  Details of the experimental setup can be found in \cite{Pyka2013b}.

In this laboratory system we trap crystals of 1 to 50 $^{172}$Yb$^{+}$ ions and observe three different phases.  Depending on the ratio of the radial to axial confinement, given by the ratio  $\nu_r/\nu_z$ of the secular frequencies at which the ions oscillate in the trap, the Coulomb crystal takes on different forms, as shown in Figure \ref{fig:phases}. 
When the radial confinement is weakened below a critical value so that $\nu_r/\nu_z<0.73 \times N^{0.86}$ where $N$ is the number of ions \cite{Steane1997}, the crystal undergoes the phase transition from a one- to a two-dimensional crystal. 
In the case of harmonic axial confinement, the phase transition spreads through the crystal from the center, where the charge density is highest.  The crystal then stabilizes in a two-dimensional zigzag geometry.  As the ratio is further decreased, the ions at the center of the trap are further pushed out into 3 dimensions, forming a helix.

\begin{figure} [t]
   \centering
  \includegraphics[scale=0.385]{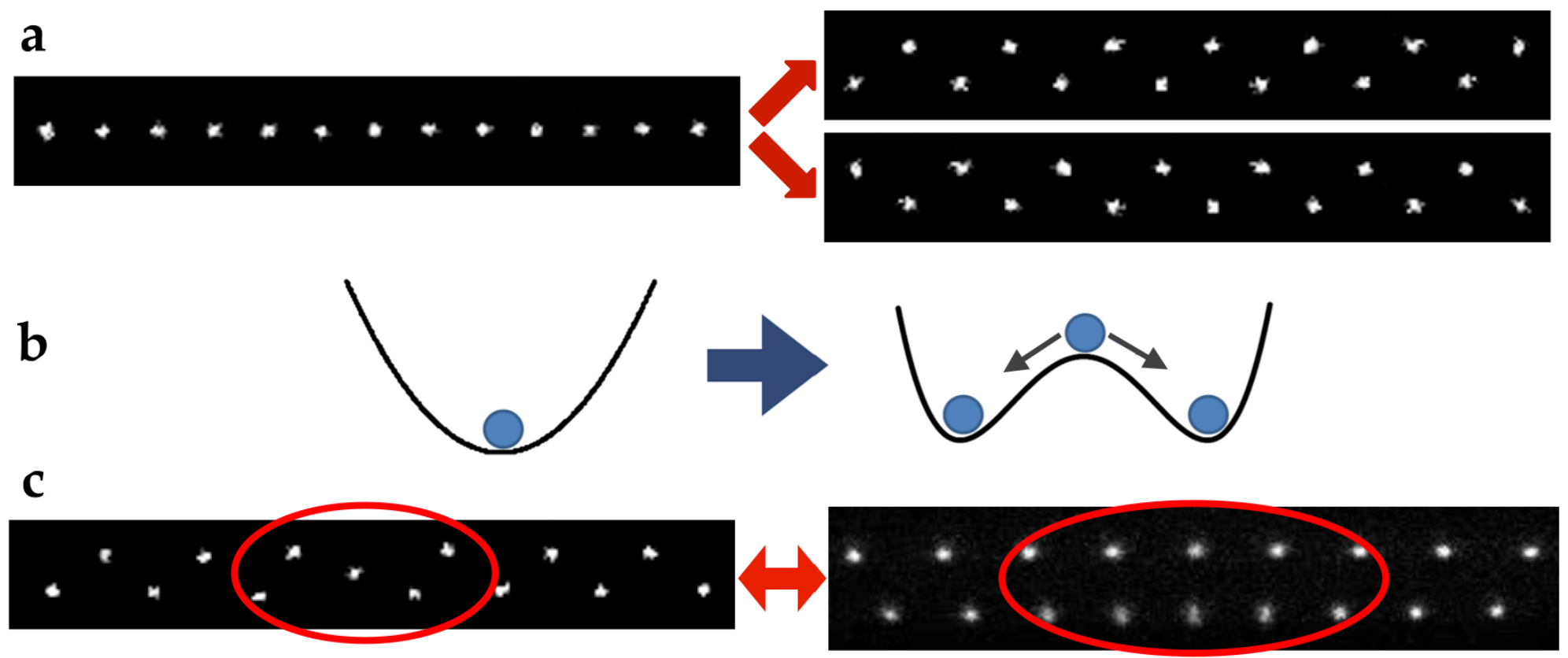} 
   \caption{Defect formation. (a) Symmetry-breaking phase transition.  The ions take on one of two possible zigzag configurations when the radial confinement is reduced and the rotational symmetry is broken. (b) The radially confining potential well becomes a double-well at the phase transition, and the ions must select one of the new minima. (c) Two types of kinks form in our system between incompatible zigzag domains; the odd kink (left) or the extended kink (right).  Switching between these configurations is accomplished by changing the ratio $\nu_r/\nu_z$. } 
	\label{fig:symmetry-breaking}
\end{figure}

We focus on the transition from linear to zigzag, which is of second order \cite{Fishman2008}.  Here the symmetry of the linear crystal is broken 
and the ions must choose between two possible zigzag states when the single-well radial potential becomes a double well (see Figure \ref{fig:symmetry-breaking}a,b). When this transition is crossed non-adiabatically, topological defects can form in the crystal with a certain probability when different domains of the crystal choose incompatible states (see Figure \ref{fig:symmetry-breaking}c). 
The defects, or kinks, are realized at the border between these incompatible domains.  We create defects by quenching the radial confinement while holding the axial confinement constant. The quenches must be fast on a timescale given by the axial secular frequency as $1/\nu_z$, which determines the speed of sound in the crystal.  This corresponds to ramp times on the order of 10 to 100 microseconds.

\paragraph{Scaling of defect density with quench rate}

\begin{figure} [t]
   \centering
  \includegraphics[scale=0.41]{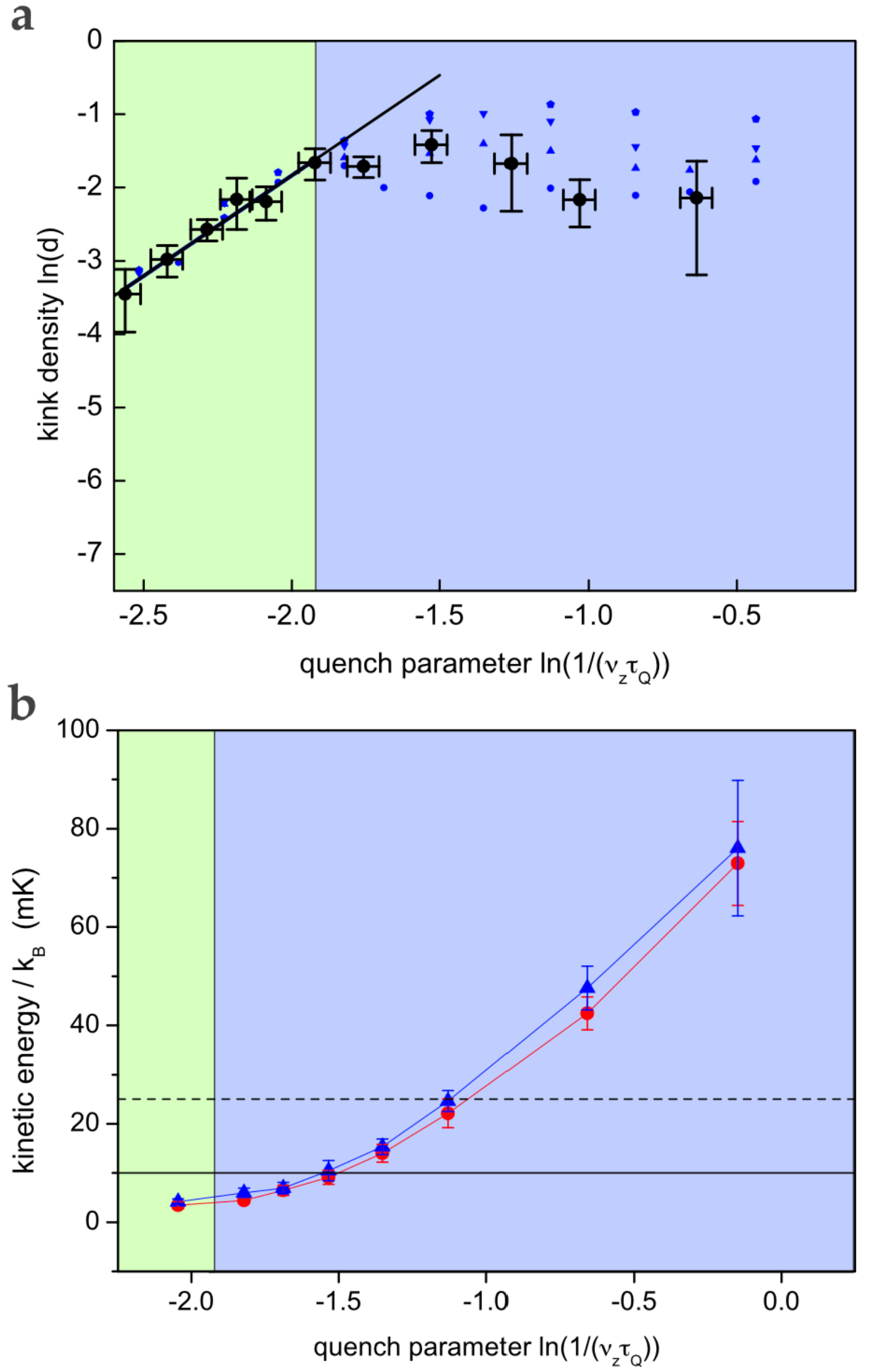} 
   \caption{ (a) Defect formation vs. quench parameter. The scaling is fitted only in the green region which is free of losses.  The black fitted line gives the measured scaling.  In the blue region, losses become apparent in the experimental data (black points), verified by numerical simulations (blue symbols) for experimental parameters. (b) Peak kinetic energy after the quench as a function of quench parameter ($1/\nu_z \tau_Q$ where $\tau_Q$ is the length in time of the quench).  The solid line indicates the depth of the odd kink PN central barrier (see Figure \ref{fig:PNpot}).  The dashed line indicates the depth of the extended kink PN barrier.  When the kinetic energy of a kink is close to or above these boundaries, losses are observed.  The red and blue symbols indicate different laser cooling parameters.
   }
	\label{fig:lossesKZM}
\end{figure}

According to the Kibble-Zurek mechanism (KZM) \cite{Kibble1980,Zurek1985}, the density of defects created while crossing the phase transition should follow a power-law scaling with varying quench rate (see \cite{delCampo2014} for a recent review).   The same scaling is expected for any system in the same universality class undergoing a second-order phase transition and is determined by the critical exponents in the Ginzburg-Landau theory \cite{Fishman2008,delCampo2010}.  We observe such a scaling in our system, however
our system has two main characteristics that deviate from the standard KZM theory: first, our system, which uses a harmonic trap, is not homogeneous, and second, the crystal is finite in size.  Accounting for these characteristics requires modifications to the theory and the expected scaling as explained in detail in \cite{Pyka2013a}. A different mathematical description of this that is rigorous without appealing to physical principles has been recently put forward \cite{Nikoghosyan2013}.  Furthermore, defect losses discussed in a later section on kink stability limit our ability to measure the number of defects created during the phase transition.  In particular, at high quench rates, a large amount of kinetic energy is introduced into the system due to the nonadiabatic quench, leading to losses.  Thus we can make reliable measurements only within a specific range of quench rates where losses do not artificially influence the results.  Figure \ref{fig:lossesKZM}a shows the observed kink density vs. quench rate, with the scaling fitted only in the green region, where we could verify using numerical simulations that there are no such losses.  In Figure \ref{fig:lossesKZM}b, the kinetic energy that is introduced into the system by the quench is given as a function of quench rate. 
These numerical results have been obtained from molecular dynamics simulations which account for the trapping and laser cooling dynamics in the time-averaged ponderomotive trap potential \cite{Pyka2013a}.

\paragraph{Types of defects}  

The type of defect which is stable for a certain ion configuration depends on the value of $\nu_r/\nu_z$, and as this ratio is changed, the configurations of both the crystal and the stable kinks can be transformed from one type to another (see Figure \ref{fig:symmetry-breaking}c).  By varying the ratio of the two different radial trap axes $\nu_{r_1}/\nu_{r_2}$, even more types of kinks can be realized, such as quasi-3d kinks \cite{Mielenz2013}.  The different families of kinks that can be produced in the dual parameter space has recently been theoretically investigated \cite{Landa2013}.

\begin{figure} 
   \centering
  \includegraphics[scale=0.385]{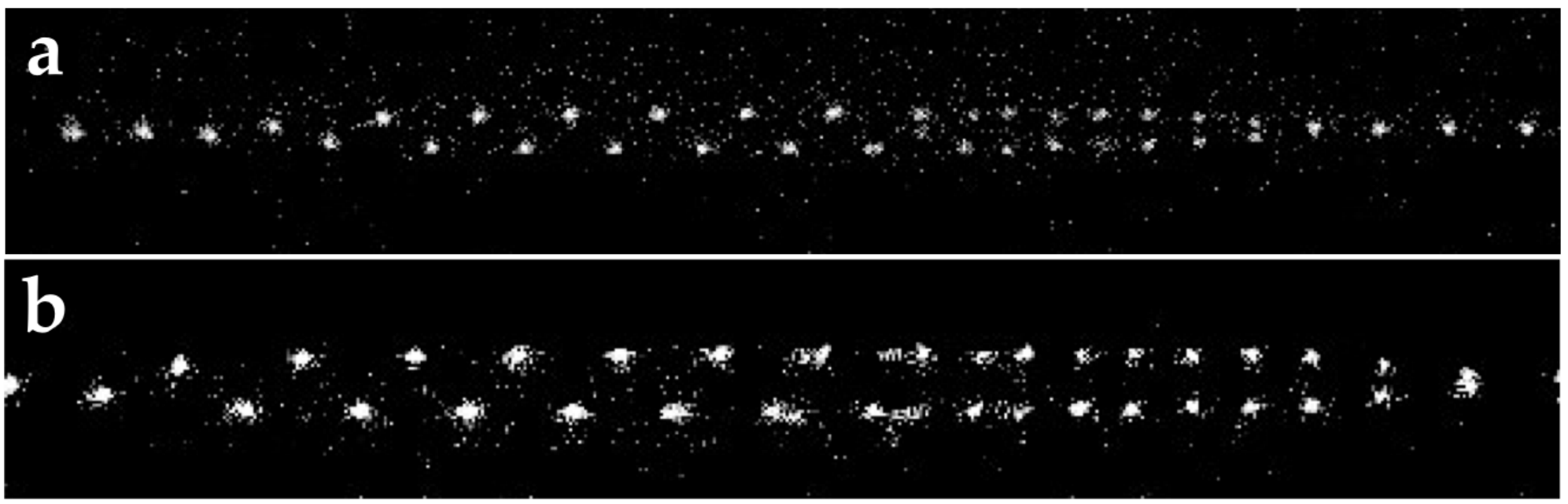} 
   \caption{ Mode of motion of the two types of kinks, as captured during a long camera exposure. (a) The odd kink leaves the crystal through radial motion of the comprising ions. (b) The extended kink moves out of the crystal through axial motion of the two separated rows. }
	\label{fig:motion}
\end{figure}

During a radial quench in our system, when the crystal first stabilizes into a zigzag, a defect takes the form of an ``odd" kink (Figure \ref{fig:symmetry-breaking}c, left), where one ion is located at the center between the two rows forming the zigzag.  In this formation, mobility of the defect occurs through radial movement of the ions (Figure \ref{fig:motion}a).  When the radial confinement is ramped to lower values, the defect transforms from odd to the ``extended" formation (Figure \ref{fig:symmetry-breaking}c, right), where the kink center is identified by ions from the upper and lower rows being located one on top of the other.  In this configuration the kink can be transported through the crystal by axial motion of the ions, since the two rows of the zigzag are now more separated and can move almost independently (Figure \ref{fig:motion}b).  At the intermediate stage between these formations, the kink can behave as either an odd or extended kink, and the mode of motion involves both axial and radial movement of individual ions.

The topological defects discussed here can be mathematically described using discrete soliton models \cite{Partner2013, Braun2004}.  They behave as quasiparticles having many of the same physical properties as solitons, e.g. effective mass, eigenfrequencies, and modes of motion, and their dynamics can be described by calculating the Peierls-Nabarro potential \cite{Boesch1989}.  

\paragraph{Defect stability} \label{losses}

\begin{figure} [t]
   \centering
  \includegraphics[scale=0.55]{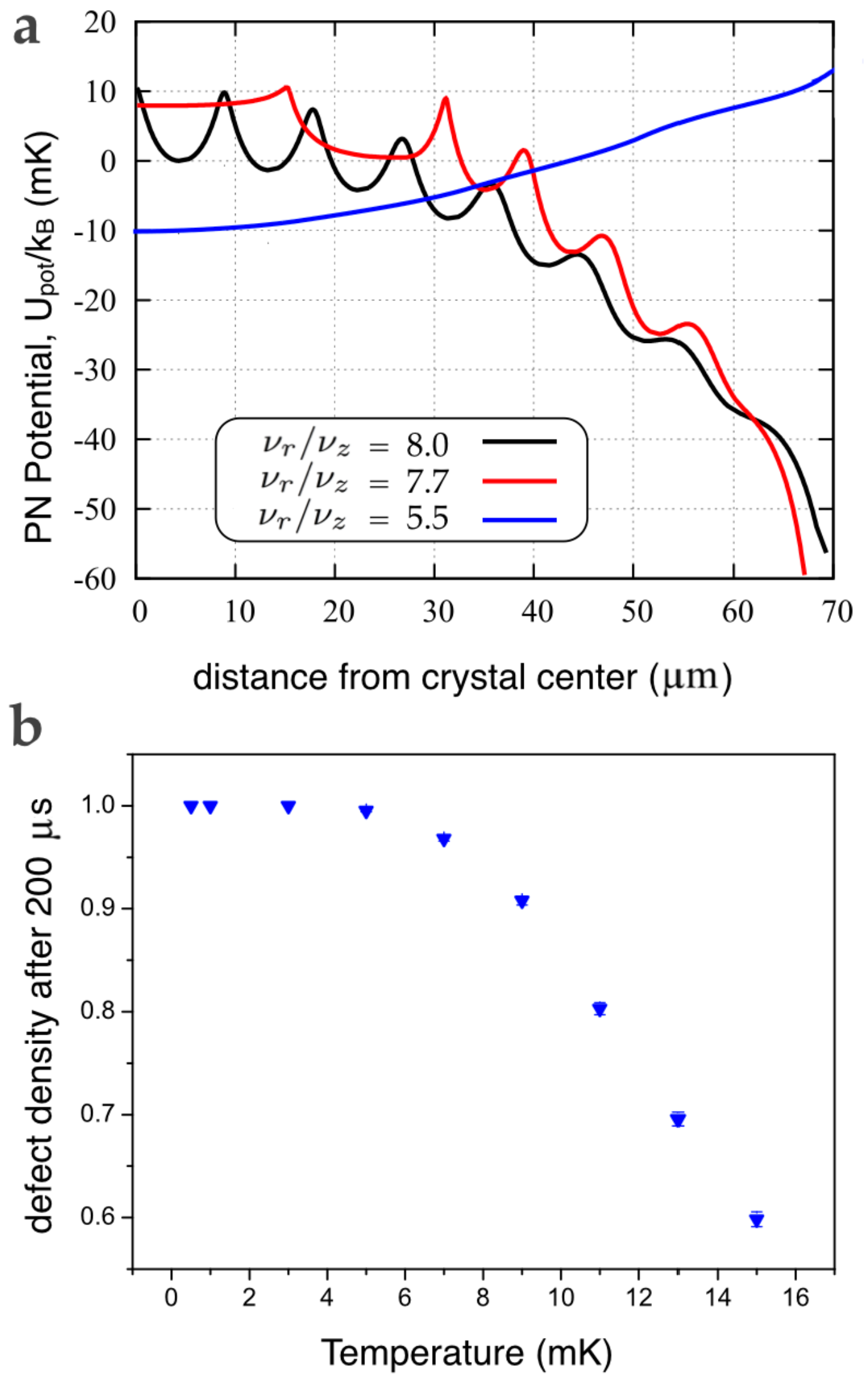} 
   \caption{(a) Peierls-Nabarro potentials as a function of kink position for different values of $\nu_r/\nu_z$, and thus for different kink configurations: odd (black), intermediate (red), and extended (blue). The absolute energy value for the different configurations is arbitrary. (b) Probability that a prepared extended kink remains in the crystal after 200 $\mu$s with increasing temperature.  
   }
	\label{fig:PNpot}
\end{figure}

Structural defects are topologically protected for a theoretical, infinite crystal.  However, due to the finite nature of our system, they can still leave the crystal through its ends.  Understanding losses of kinks that are created at the phase transition, which occur on a microsecond timescale, is critical for interpreting experimental data, which is measured on the scale of milliseconds.  Several factors can lead to losses, and these losses must be understood through simulations and by evaluating the potential experienced by the defect.  This is the Peierls-Nabarro (PN) potential which changes during the quench as illustrated in Figure \ref{fig:PNpot}a for a model using our experimental parameters.  It is calculated by moving a simulated kink through the crystal, and evaluating the total potential energy of the configuration with the kink at each point \cite{Partner2013}.  

During a quench across the phase transition, when the kink first appears, the PN potential has a global maximum at the center of the crystal, 
encouraging a kink to move to the end of the trap where it is lost.  However, as the ratio $\nu_r/\nu_z$ reaches the regime of an odd kink, local minima form where an odd kink can be stabilized, but the global maximum persists which allows odd kinks with a certain energy to be lost.  For example, an odd kink with a kinetic energy ($E_{\rm{kin}}/k_B$) of 10 mK can escape the deepest wells and leave the crystal; the thermal spread of kink energies causes losses to occur for considerably lower system temperatures.
As the radial confinement is lowered toward the extended kink regime, the shape of the PN potential evolves through the intermediate regime to become a globally confining potential that is realized for extended kinks.   For our experimental parameters shown in Figure \ref{fig:PNpot}a, the energy barrier for an extended kink to leave the crystal is already more than a factor of 2 higher than for the odd kink case. Due to this confining PN potential, extended kinks provide a signature of the number of kinks created during a quench that is detectable at longer time scales.   However, due to the global minimum of the extended kink potential, one can only observe single extended kinks since pairs of these kinks will be attracted to the center of the crystal and annihilate.  Ultimately, the lifetime of a trapped kink in the ion crystal is limited by collisions from background gases, which heat up the entire crystal and destroy any previous structure, including defects. 

The kinetic energy of a kink is mainly influenced by the rate of the quench. Figure \ref{fig:lossesKZM}b illustrates the influence of quench rate on kinetic energy for simulations performed using our experimental parameters.  Additional energy produced by faster quenches increases the probability of a kink to cross potential barriers and be lost before detection.  
The horizontal lines in Figure \ref{fig:lossesKZM}b illustrate the maximum kinetic energy that can be trapped within the PN barriers.  These values are consistent with observed losses in the experiment (Figure \ref{fig:lossesKZM}a).  Thus, the range of quench rates that support stable trapping is determined largely by the PN potential barriers.  By engineering the PN potential using trap parameters, as well as mass defects and electric fields (discussed next), it will be possible to extend the range in which measurements are unaffected by losses.  

We have also performed simulations of losses depending on the temperature in the system, shown in figure \ref{fig:PNpot}b.  The probability for a prepared (extended) defect to remain in the crystal for 200 $\mu$s is plotted as a function of system temperature.  The corresponding extended kink PN barrier (blue) in Figure  \ref{fig:PNpot}a is about 20 mK.  This simulation shows that losses start to occur at temperatures as low as 6 mK, corresponding to losses from the high-energy tail of the Boltzmann distribution.

\begin{figure} [t]
   \centering
  \includegraphics[scale=0.385]{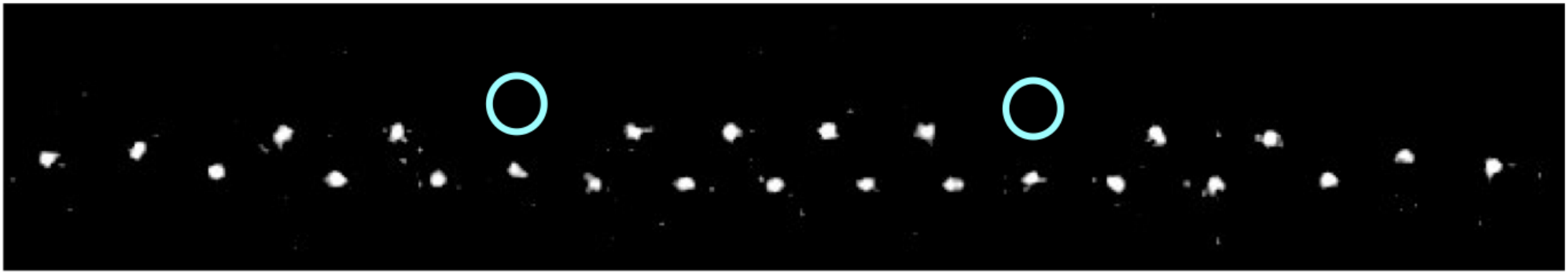} 
   \caption{Two extended kinks trapped next to each other with the help of heavy mass impurities. The open circles indicate the location of the two heavier ions, which do not fluoresce.   
   }
	\label{fig:twokinks}
\end{figure}

\paragraph{Kink dynamics in the presence of mass defects and electric fields} Understanding kink formation and stability lays the groundwork for working with kinks, but adding ingredients such as electric fields or different masses within the crystal modifies the dynamics, and can add an element of control.  We have investigated the dynamics of kinks within the PN potential when a mass defect is present in the crystal.  This can occur in our system when a collision from the background gas causes a molecule to form with one of the Yb$^{+}$ ions.  Due to the mass dependency of the radial confinement in an RF ion trap, these impurity ions alter the PN potential and can be used to manipulate kink dynamics and stability.  For example, a heavy mass defect forms an additional potential minimum that can localize a kink at its position. With two mass defects in the zigzag one can even engineer a PN potential that traps two extended kinks that do not annihilate in the same finite crystal (see Figure \ref{fig:twokinks}), which is not possible in the pure crystal case. In addition, by applying static electric fields, one can tune the influence of the mass defect on the Coulomb crystal by changing its effective mass.  Together with mass defects, electric fields can break the symmetry of the crystal in a controlled way, allowing deterministic creation of structural defects.   We discuss these effects in more detail in another work focused on defect dynamics \cite{Partner2013}.

\section{Conclusions and Outlook}

We have created and stably trapped topological defects in ion Coulomb crystals.  Implementing high resolution laser spectroscopy will enable studies of these quasiparticles and their localized vibrational mode structure \cite{Landa2010}. In the future, implementing ground state cooling of the internal modes of a kink will allow investigation of quantum effects and exploration of kink-kink interactions.  Finally, the Aubrey transition of pinned particles in a periodic potential can be studied for the incommensurate case using extended kinks in ion Coulomb crystals \cite{sharma1984, Peyrard1983}.


\section*{Acknowledgements}
This work was supported by DFG through QUEST and by the EU Integrating Project SIQS, the EU STREP EQUAM and an Alexander von Humboldt Professorship.


\bibliography{Mehlstaeubler_bib}

\end{document}